\documentclass[runningheads]{llncs}
\usepackage{graphicx}
\usepackage{amsmath}
\usepackage{amsfonts}
\usepackage{array}
\usepackage{multirow}

\newcommand{\argmin}{\mathop{\mathrm{argmin}}}
\newcommand{\etal}{\textit{et al}.}
\newcommand{\ie}{\textit{i}.\textit{e}.}
\newcommand{\eg}{\textit{e}.\textit{g}.}

\begin{document}
\title{SVoRT: Iterative Transformer for Slice-to-Volume Registration in Fetal Brain MRI}
\titlerunning{SVoRT}
%

\author{Junshen Xu\inst{1} \and
Daniel Moyer\inst{2} \and
P. Ellen Grant\inst{3,4} \and
Polina Golland\inst{1,2} \and
Juan Eugenio Iglesias\inst{2,4,5,6}
Elfar Adalsteinsson\inst{1,7}
}

%
\authorrunning{J. Xu et al.}

\institute{Department of Electrical Engineering and Computer Science, MIT, \\ Cambridge, MA, USA \\\email{junshen@mit.edu} \and
Computer Science and Artificial Intelligence Laboratory, MIT, \\
Cambridge, MA, USA \and
Fetal-Neonatal Neuroimaging and Developmental Science Center, \\ Boston Children’s Hospital, Boston, MA, USA \and
Harvard Medical School, Boston, MA, USA \and
Centre for Medical Image Computing, Department of Medical Physics and Biomedical Engineering, University College London, UK \and
Athinoula A. Martinos Center for Biomedical Imaging, Massachusetts General Hospital and Harvard Medical School, Boston, USA \and
Institute for Medical Engineering and Science, MIT, Cambridge, MA, USA}

\maketitle              
\begin{abstract}
Volumetric reconstruction of fetal brains from multiple stacks of MR slices, acquired in the presence of almost unpredictable and often severe subject motion, is a challenging task that is highly sensitive to the initialization of slice-to-volume transformations. We propose a novel slice-to-volume registration method using Transformers trained on synthetically transformed data, which model multiple stacks of MR slices as a sequence. With the attention mechanism, our model automatically detects the relevance between slices and predicts the transformation of one slice using information from other slices. We also estimate the underlying 3D volume to assist slice-to-volume registration and update the volume and transformations alternately to improve accuracy. Results on synthetic data show that our method achieves lower registration error and better reconstruction quality compared with existing state-of-the-art methods. Experiments with real-world MRI data are also performed to demonstrate the ability of the proposed model to improve the quality of 3D reconstruction under severe fetal motion.

\end{abstract}

\section{Introduction}
Volumetric reconstruction of fetal brain from multiple motion-corrupted stacks of MR slices is an important tool in studying fetal brain development~\cite{gholipour2017normative,benkarim2020novel,vasung2019exploring}. 
Due to rapid and random fetal motion, fetal MRI are limited to fast acquisition techniques, such as the single-shot T2 weighted (SST2W) imaging that freezes in-plane motion. 
Even with such fast 2D sequences, fetal MRI is still vulnerable to inter-slice motion artifacts~\cite{malamateniou2013motion}, leading to misalignment of slices in a stack.
Moreover, the delay between slices due to safety constraints on specific absorption rate (SAR)~\cite{krishnamurthy2015mr} further worsen the situation.
Therefore, Slice-to-Volume Registration (SVR) prior to 3D reconstruction of fetal brain is necessary. 
Manual SVR is usually infeasible in practice due to the magnitude of data involved.
Although optimization-based SVR methods have successfully applied to 3D reconstruction of fetal brain~\cite{kainz2015fast,kuklisova2012reconstruction,gholipour2010robust}, coarse alignment of slices is required to initialize the algorithm and the quality of reconstructed volume is highly dependent on the initial alignment.
Hence, an automatic and accurate method for estimating slice transformations is crucial to 3D reconstruction of fetal brain.

In an attempt to speed up SVR of fetal MRI and improve its capture range, deep learning methods~\cite{hou20183,salehi2018real} have been proposed to predict rigid transformations of MR slices using Convolution Neural Networks (CNNs), which share similarity with camera pose prediction in computer vision~\cite{hou2018computing,kendall2015posenet}. Pei~\etal~\cite{pei2020anatomy} proposed a multi-task network to exploit semantic information in fetal brain anatomy which, however, requires annotations of segmentation maps. Moreover, these approaches process each slice independently, ignoring the dependencies between slices. Singh~\etal~\cite{singh2020deep} proposed a recurrent network to predict inter-slice motion in fetal MRI. In SVR of fetal ultrasound, Yeung~\etal~\cite{yeung2021learning} tried to predict the 3D location of multiple slices simultaneously with an attention CNN.

Recently, Transformer models~\cite{vaswani2017attention} and their variants have achieved astounding results in various fields~\cite{dosovitskiy2020image,devlin2019bert}.
The concept behind Transformers is to dynamically highlight the relevant features in input sequences with the self-attention mechanism, which demonstrates great capability of modeling long-distance dependencies and capturing global context. 
In SVR of fetal MRI, multiple stacks of slices are provided as inputs, which can also be modeled as a sequence of images.
Multi-view information from stacks of slices with different orientations can be processed jointly to assist the SVR task.

Here, we propose a Slice-to-Volume Registration Transformer (SVoRT) to map multiple stacks of fetal MR slices into a canonical 3D space and to further initialize SVR and 3D reconstruction. 
As such, we present the following contributions:
1) We propose a Transformer-based network that models multiple stacks of slices acquired in one scan as a sequence of images and predicts rigid transformations of all the slices simultaneously by sharing information across the slices.
2) The model also estimates the underlying 3D volume to provide context for localizing slices in 3D space.
3) In the proposed model, slice transformations are updated in an iterative manner to progressively improve accuracy.

\section{Methods}

\begin{figure}[t]
\centering
\includegraphics[width=0.9\textwidth]{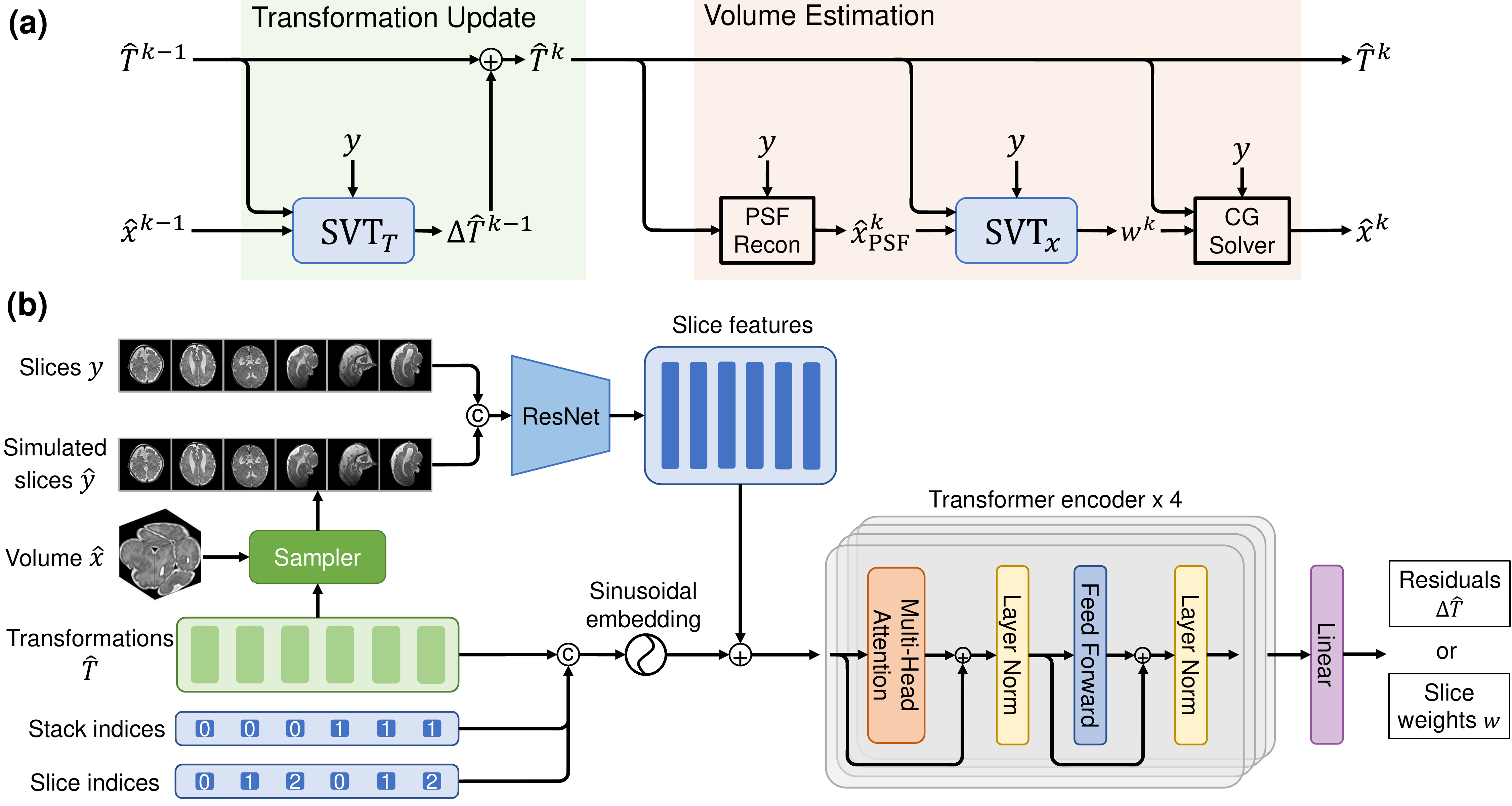}
\caption{(a) The $k$-th iteration of SVoRT. (b) The architecture of SVT.} \label{fig:method}
\end{figure}

Given $n$ acquired slices of a scan, $y=[y_1,\dots,y_n]$, the goal of SVoRT is to estimate the corresponding transformations $T=[T_1, \dots, T_n]$,~\ie, rotations and translations of acquisition planes, in a 3D canonical atlas space. 
However, unlike SVR problems~\cite{esteban2019towards,gillies2017real} where a 3D volume exists as a reference for matching 2D slices, high quality 3D references are usually unavailable in SVR of fetal MR due to fetal motion. 
Therefore, instead of predicting the transformations alone, we also estimate the underlying volume $x$ from the input slices, so that the estimated volume $\hat{x}$ can provide 3D context to improve the accuracy of predicted transformations.
In SVoRT, the estimated transformation $\hat{T}$ and the estimated volume $\hat{x}$ are updated alternately, generating a series of estimates, $(\hat{T}^0, \hat{x}^0), \dots, (\hat{T}^K, \hat{x}^K)$, where $\hat{T}^0$ and $\hat{x}^0$ are the initial guesses, and $K$ is the number of iterations.
The estimated transformations of the last iteration $\hat{T}^K$ is used as the output of the model. Fig.~\ref{fig:method} (a) shows the $k$-th iteration of SVoRT, which consists of two steps: 1) computing the new transformation $\hat{T}^{k}$ given $\hat{T}^{k-1}$ and $\hat{x}^{k-1}$ from the previous iteration, and 2) estimating volume $\hat{x}^k$ based on the new transformation $\hat{T}^k$.

\subsection{Transformation Update}
At the $k$-th iteration, the transformations are updated by
$\hat{T}^{k}=\hat{T}^{k-1}+\Delta\hat{T}^{k}$. We propose a submodule named Slice-Volume Transformer (SVT) to regress the residual $\Delta\hat{T}^k$ given the set of input slice and the estimates from the previous iteration, $\Delta\hat{T}^k=\text{SVT}_T^k(y,\hat{T}^{k-1},\hat{x}^{k-1})$.
SVT, whose architecture is shown in Fig.~\ref{fig:method} (b), aims to jointly extract features from stacks of slices and a 3D volume. 

To relate the volume $\hat{x}^{k-1}$ to the set of slices $y$ with estimated transformations $\hat{T}^{k-1}$, SVT simulates slices from the volume following the forward model, $\hat{y}_i=DB\hat{T}_i^{k-1}\hat{x}^{k-1}$, $i=1,\dots,n$, where $D$ and $B$ are the operators for slice sampling and Point-Spread-Function (PSF) blurring respectively. 
The simulated slices $\hat{y}$ provide views of the estimated volume $\hat{x}$ at the estimated slice locations $\hat{T}^{k-1}$. 
The difference between $\hat{y}$ and the original slices $y$ can be used as a proxy indicator of registration accuracy and guide models to update the estimated transformations.
To this end, we concatenate $\hat{y}$ and $y$, and use a ResNet~\cite{he2016deep} to extract features $X^\text{slice}\in\mathbb{R}^{n\times d}$ from slices, where $d$ is the number of features. 

In addition to the image content, information about the position of the slice in the sequence and the estimated location in 3D space is injected for Transformers to encode spatial correlation of the input sequence, ~\eg, adjacent slices in the same stacks are usually highly correlated, while stacks with different orientations provide complementary information. Each slice in the input sequence is associated with two indices, the index of the stack that the slice belongs to, and the index of the slice in the stack. Positional embeddings $X^\text{pos}\in\mathbb{R}^{n\times d}$ are generated from the current estimated transformation $\hat{T}^{k-1}$, the stack indexes, and the slice indexes using sinusoidal functions~\cite{vaswani2017attention}.

The slice features and the corresponding positional embeddings are added and provided to a Transformer with four encoders~\cite{vaswani2017attention}. Each Transformer encoder consists of a multi-head attention module, a feed forward network, and two layer normalization~\cite{ba2016layer}. Let $X=X^\text{slices} + X^\text{pos}$ be the input matrix. The multi-head attention module first projects $X$ into three different spaces, $Q_j=XW_j^Q, K_i=XW_j^K, V_j=XW_j^V$, $j=1,\dots,h$, where weights $W_j^Q, W_j^K, W_j^V\in\mathbb{R}^{d\times(d/h)}$ and $h$ is the number of heads. Then, each head computes the output as $Y_j=\text{softmax}(Q_j K_j^T/\sqrt{d})V_j$, where the softmax function is applied to each row. 
The outputs from all heads are aggregated and projected with a matrix $W^O\in\mathbb{R}^{d\times d}$,~\ie, $Y=[Y_1,...,Y_h]W^O$.
The feed forward network is a fully connected network used to extract deeper features.
At the end, a linear layer is applied to regress the residual transformations from the output of Transformer. 

\subsection{Volume Estimation}
The next step is to compute the new estimate of volume $\hat{x}^{k}$ based on the updated transformations $\hat{T}^{k}$. 
One of the available methods is the PSF reconstruction~\cite{kainz2015fast}, which aligns the slices in 3D space based on transformations $\hat{T}^{k}$ and interpolates the volume with the PSF kernel. 
However, there are two disadvantages of this approach. First, it over smooths the reconstructed volume and leads to a loss of image detail. Second, it fails to exclude slices with large transformation error during reconstruction, resulting in artifacts in the reconstructed volume. 

To address these problems, we use another SVT in the volume estimation step to predict weights of slices, $w=[w_1, \dots, w_n]$, where $w_i\in[0,1]$ represents the image quality of the $i$-th slice. The SVT here shares the same architecture as the one in the transformation update step, but has different inputs. Specifically, in the inputs to SVT, $\hat{T}^{k-1}$ and $\hat{x}^{k-1}$ are replaced with the updated transformations $\hat{T}^k$ and the PSF reconstruction result $\hat{x}_\text{PSF}$ respectively, $w^k=\text{SVT}_x^k(\hat{T}^k, \hat{x}_\text{PSF}^k,y)$, where we denote the SVT in volume estimation as $\text{SVT}_x^k$.  To compute the new estimated volume $\hat{x}^k$, we solve an inverse problem,
\begin{equation}
    \hat{x}^{k}=\argmin_{x}\sum_{i=1}^n w_i^{k}\|DB\hat{T}_i^{k}x-y_i\|_2^2,
    \label{eqn:vol}
\end{equation}
where the weights help exclude outliers during volume estimation. Since the closed form solution involves inverting
a very large matrix, we instead employ a conjugate gradient (CG) method to compute $\hat{x}^k$. Note that all operations (matrix multiplication, addition and scalar division) in CG are differentiable, so the gradient with respect to $w$ can be computed via automatic differentiation.

\subsection{Training}
{\bf Data:}
Supervised learning of SVoRT requires the ground truth transformation of each slice. However, annotating the 3D location of a 2D MR slice is very challenging. 
Instead, we artificially sample 2D slices as training data from high quality MR volumes of fetal brain reconstructed from data with little fetal motion. The orientations of stack,~\ie, the normal vector and in-plane rotation, are randomly sampled as in~\cite{hou20183} so that the dataset captures a wide range of rigid transformations. To Bridge the gap between the synthetic data and real MR scans and improve the generalization capability of networks, we also adopt various data augmentation and MR artifact simulation techniques~\cite{iglesias2021joint}, including image noise, bias field, signal void artifacts, and contrast jitter.

{\bf Representations of transformations:} Various representations are available for the describing the location of a plane in 3D space. For example, the Euler angles, the affine matrix, or the Cartesian coordinates of 3 points, called anchor points, within the plane. Previous works have demonstrated that the anchor point representation yields the highest accuracy for deep learning based SVR methods. Following~\cite{hou20183}, we use the center, the bottom right and left corners of a plane as the anchor points to define the rigid transformation.

{\bf Loss functions:}
During training, we apply the L2 loss between the predicted and target anchor points for transformation prediction, 
\begin{equation*}
\mathcal{L}_T^k=\| \hat{P}_1^k-P_1\|_2^2+\|\hat{P}_2^k-P_2\|_2^2+\|\hat{P}_3^k-P_3\|_2^2,    
\end{equation*}
where $\hat{P}_1^k,\hat{P}_2^k,\hat{P}_3^k$ are the predicted coordinates of the three anchor points in the $k$-th iteration, and $P_1,P_2,P_3$ are the ground truth locations. As for volume estimation, the L1 loss between the $k$-th estimated volume and the target volume is used, $\mathcal{L}_x^k=\|\hat{x}^k -x\|_1$. The total loss $\mathcal{L}$ is the sum of the losses in all iterations, $\mathcal{L}=\sum_{k=1}^K\mathcal{L}_T^k+\lambda \sum_{k=1}^K\mathcal{L}_x^k,$
where $\lambda$ is a weight determining the relative contribution of the L1 loss.

\section{Experiments and Results}
\subsection{Experiment Setup}
We evaluate the models on the FeTA dataset~\cite{payette2021automatic}, which consists of 80 T2-weighted fetal brain volumes with gestational age between 20 and 35 weeks. The dataset is split into training (68 volumes) and test (12 volumes) sets. The volumes are registered to a fetal brain atlas~\cite{gholipour2017normative}, and resampled to 0.8 mm isotropic. We simulate 2D slices with resolution of 1 mm $\times$ 1 mm, slice thickness between 2.5 and 3.5 mm, and size of 128 $\times$ 128. Each training sample consists of 3 image stacks in random orientations and each stack has 15 to 30 slices. Fetal brain motion is simulated as in~\cite{xu2021stress} to perturb the transformations of slices. In the process of training, random samples are generated on the fly, while for testing, 4 different samples are generated for each test subject, resulting in 48 test cases. 

To demonstrate SVoRT can generalize well to real-world data and help initialize SVR for cases with severe fetal motion, we test the trained models with data acquired in real fetal MRI scans. Scans were conducted at Boston Children's Hospital. MRI data were acquired using the HASTE sequence~\cite{gagoski2021automated} with slice thickness of 2 mm, resolution of 1 mm $\times$ 1 mm, size of 256 $\times$ 256, $\text{TE}=119\text{ ms}$, and $\text{TR}=1.6\text{ s}$. The real MR dataset not only has different contrast, but also undergoes more realistic artifacts and fetal motion compared to the synthetic data. 
In the experiments, SVRnet~\cite{hou20183} and PlaneInVol~\cite{yeung2021learning} ared used as baselines. SVRnet predicts the transformation of each slice independently with a VGGNet~\cite{simonyan2014very}, while PlaneInVol uses an attention CNN which compares pairs of slices, and learns to map a set of slices to 3D space. For SVoRT, we set the initial estimates $\hat{T}^{0}$ and $\hat{x}^0$ to identity transformation and zero respectively and set $\lambda=10^3$ and $K=3$. All neural networks are implemented with PyTorch~\cite{paszke2017automatic} and trained on a 
Nvidia Tesla V100 GPU with an Adam optimizer~\cite{kingma2017adam} for $2\times10^5$ iterations. We used an initial learning rate of $2\times 10^{-4}$ which linearly decayed to zero.
The reference implementation for SVoRT is available on GitHub\footnote{https://github.com/daviddmc/SVoRT}.

\subsection{Simulated Data}
To evaluate the accuracy of the predicted transformation for different models, we use the Euclidean Distance (ED) of anchor points, and the Geodesic Distance (GD) in $SO(3)$: $\text{ED}=\frac{1}{3}\sum_{j=1}^3\|\hat{P}_j-P_j\|_2$ and $\text{GD}=\text{arccos}(\frac{\text{Tr}(R)-1}{2})$, 
where $R$ is the rotation matrix from the predicted plane to the target, representing the rotation error. 
We also extract slices from the ground truth volume $x$ according to the estimated transformations $\hat{T}$ and compare them to the original slices $y$. Comparison is performed via Peak Signal-to-Noise Ratio (PSNR) and Structural Similarity (SSIM)~\cite{wang2004image}.
To further examine the model, we use the predicted transformations to initialize a 3D reconstruction algorithm~\cite{kainz2015fast} and compute PSNR and SSIM between the reconstructed volumes and the targets.

\newsavebox\CBox
\def\textBF#1{\sbox\CBox{#1}\resizebox{\wd\CBox}{\ht\CBox}{\textbf{#1}}}
\newcolumntype{x}[1]{>{\centering\arraybackslash\hspace{0pt}}p{#1}}
\newcommand{\tabcol}{1.668cm}
\def\arraystretch{1.0}
\begin{table}
\caption{Mean values of quantitative metrics for different models (standard deviation in parentheses). $\downarrow$ indicates lower values being more accurate, vice versa. The best results are highlighted in bold.}\label{tab:simulated}
\centering
\begin{tabular}{@{\extracolsep{0pt}}c|x{\tabcol}x{\tabcol}|x{\tabcol}x{\tabcol}|x{\tabcol}x{\tabcol}@{}}
\hline
\multirow{2}{*}{Method} & \multicolumn{2}{c|}{Transformation} & \multicolumn{2}{c|}{Slice} & \multicolumn{2}{c}{Volume}\\
\cline{2-3}\cline{4-5}\cline{6-7}
 & \centering{ED(mm)$\downarrow$} & GD(rad)$\downarrow$ & PSNR$\uparrow$ & SSIM$\uparrow$ & PSNR$\uparrow$ & SSIM$\uparrow$\\
\hline
SVRnet     & 12.82 (5.69) & .256 (.150) & 20.53 (1.62) & .823 (.061) & 19.54 (1.52) & .669 (.116) \\
PlaneInVol & 12.49 (6.73) & .244 (.213) & 19.96 (1.73) & .808 (.069) & 18.98 (1.62) & .615 (.139) \\
\hline
SVoRT & \textBF{4.35} (0.90) & \textBF{.074} (.017) & \textBF{25.26} (1.86) & \textBF{.916} (.034) & \textBF{23.32} (1.42) & \textBF{.858} (.037) \\
w/o PE     &  9.97 (6.28) & .194 (.179) & 21.44 (2.08) & .841 (.064) & 20.74 (1.49) & .742 (.096) \\
w/o Vol    &  5.09 (1.05) & .088 (.020) & 23.97 (1.68) & .894 (.040) & 22.89 (1.37) & .844 (.043) \\
$K=1$      &  5.99 (1.16) & .103 (.024) & 23.02 (1.67) & .876 (.047) & 22.57 (1.21) & .836 (.041) \\
$K=2$      &  5.65 (1.07) & .097 (.022) & 23.25 (1.84) & .878 (.048) & 22.64 (1.50) & .837 (.043) \\ 
\hline
\end{tabular}
\end{table}

\begin{figure}[t]
\centering
\includegraphics[width=\textwidth]{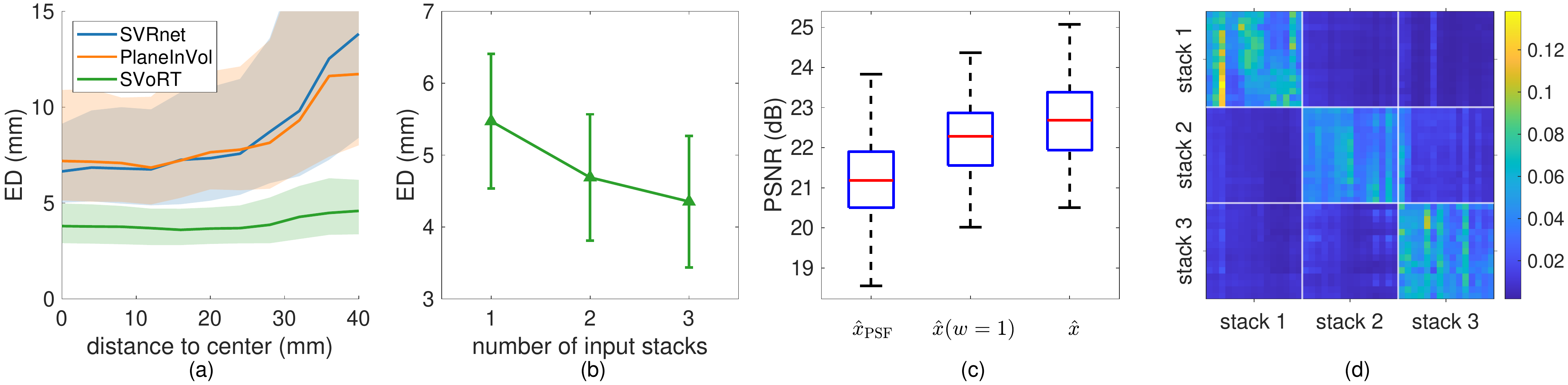}
\caption{(a) Medians of ED for slices sampled at different locations at the atlas space. Error bands indicate 25 and 75 percentiles. (b) Mean values of ED for SVoRT models with different numbers of input stacks. Error bars indicate standard deviations (c) PSNRs of different volume estimation methods. (d) An example heatmap of the self-attention weight matrix averaged over all the heads.} \label{fig:misc}
\end{figure}

Table~\ref{tab:simulated} reports the mean and standard deviations of quantitative metrics for different models on the test set of the simulated data. Our proposed method outperforms both SVRnet and PlaneInVol, which only leverage the intensity information of slices. As shown in Fig.~\ref{fig:misc} (a), the transformation errors for SVRnet and PlaneInVol increase with the distance to the center of 3D space, since the slices near the boundary of fetal brain contain little content and can be ambiguous. However, by exploiting the positional information of slice in the input sequence, SVoRT is able to register such cases better and lead to lower errors.
\begin{figure}[!h]
\centering
\includegraphics[width=0.9\textwidth]{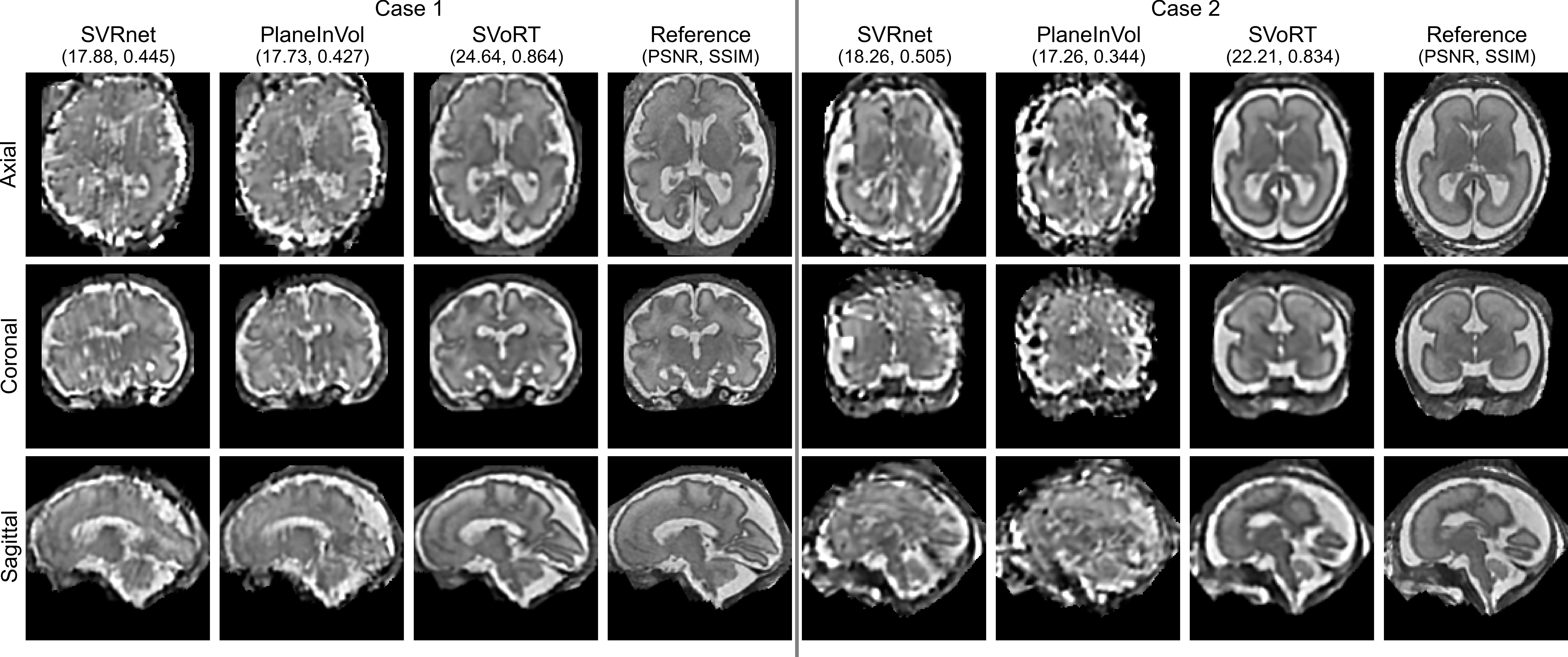}
\caption{Example reconstructed volumes and reference volumes of the test set.} \label{fig:simulated}
\end{figure}

Ablation studies are also performed by removing the positional embedding (w/o PE) and the volume estimation (w/o Vol), and using fewer iterations ($K=1,2$) in SVoRT. Results indicate that the positional embedding serves as a prior for the relative locations of slices in a stack, which facilitates the registration process. The auxiliary volume estimation improves the accuracy of transformation prediction by providing 3D context. Moreover, the iterative update enable the model to progressively refine the predictions. We test SVoRT with different numbers of input stacks (Fig.~\ref{fig:misc} (b)). With more input stacks, SVoRT receives more different views of the 3D volume, and therefore achieves lower registration error. We also compare different volume estimation methods: i) the proposed estimate $\hat{x}$, ii) the solution to the inverse problem in Eq.~(\ref{eqn:vol}) with equal weight, $\hat{x}(w=1)$, and iii) the PSF reconstruction, $\hat{x}_\text{PSF}$. As shown in Fig.~\ref{fig:misc} (c), the proposed method achieves the highest PSNR.
Fig.~\ref{fig:misc} (d) visualizes an example attention matrix generated by the last Transformer encoder.
The 3D reconstruction results in Fig.~\ref{fig:simulated} show that SVoRT also yields better perceptual quality compared with other state-of-the-art methods, in consequence of fewer slice misalignment in the initialization of SVR.


\subsection{Real Fetal MR Data}

We collect 3 orthogonal motion-corrupted stacks of MR slices from two subjects respectively.
For preprocessing, bias fields are corrected~\cite{tustison2010n4itk}, and the brain ROI is manually segmented from each slice. 
SVRnet, PlaneInVol and SVoRT are used to predict the transformations and initialize the SVR algorithm. For comparison, we also apply the SVR algorithm to the input stacks directly without deep learning based initialization (SVR only). Note that the results of ``SVR only'' are reconstructed in the subject space. We further register them to the atlas space for visualization.
\begin{figure}[t]
\centering
\includegraphics[width=0.9\textwidth]{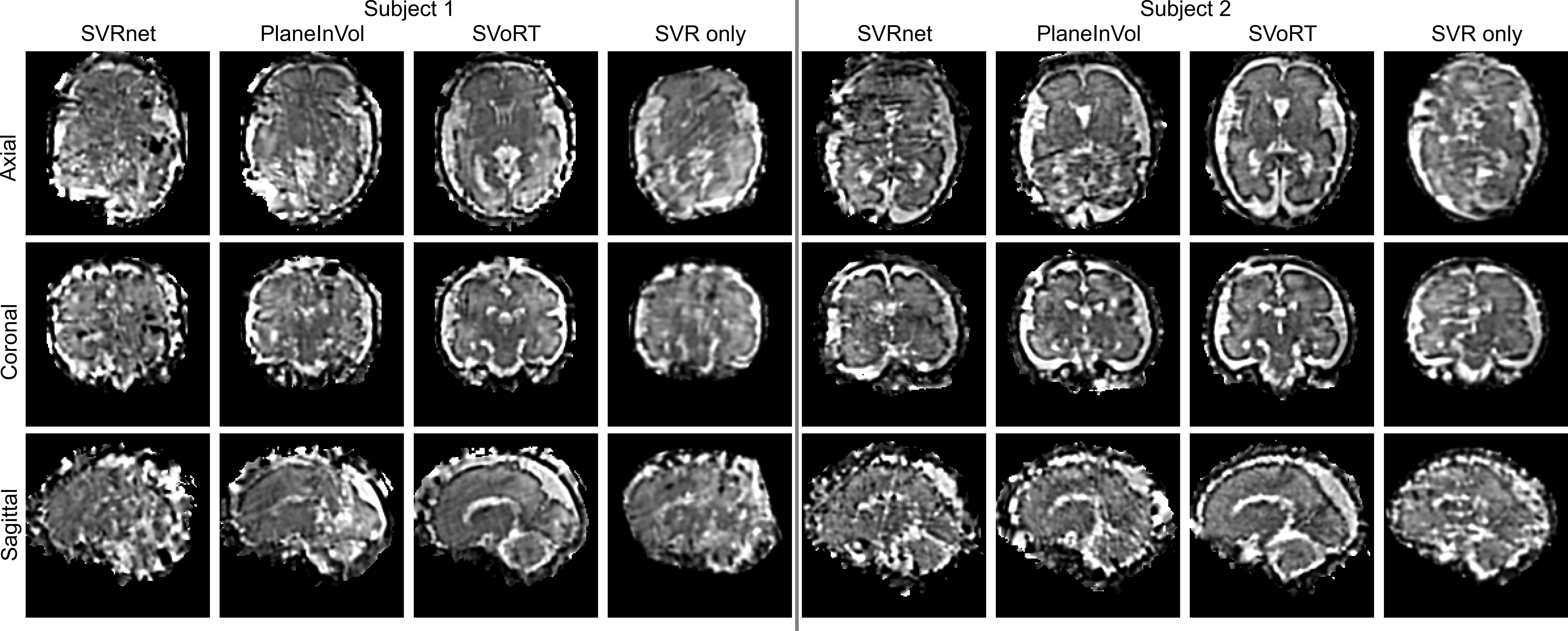}
\caption{Reconstructed volumes for different methods of real MR data.} \label{fig:real}
\end{figure}
Fig.~\ref{fig:real} shows that volumes reconstructed by SVR alone suffer from severe image artifacts due to slice misalignment caused by drastic fetal motion.
SVRnet and PlaneInVol are incapable of generalizing to real MR data and fail to provide a useful initialization for SVR. 
In comparison, the estimated transformations of SVoRT are more accurate and the corresponding reconstructed volume presents better perceptual quality.
Results indicate that SVoRT learns more robust features from synthetic data, and therefore generalizes well in the presence of real-world noise and artifacts. Moreover, the average inference time of SVoRT for each subject is 0.8 s, which is negligible compared with SVR algorithms that usually take minutes even on GPUs. SVoRT potentially enables high quality 3D reconstruction of fetal MRI in the case of severe fetal motion. 

\section{Conclusion}
In this work, we propose a novel method for slice-to-volume registration in fetal brain MRI using Transformers. By jointly processing the stacks of slices as a sequence, SVoRT registers each slice by utilizing context from other slices, resulting in lower registration error and better reconstruction quality. In addition, we introduce an auxiliary task of volume estimation and update the transformation iteratively to improve registration accuracy. Evaluations show that SVoRT learns more robust features so that, by training on simulated data, it generalizes well to data acquired in real scans. SVoRT provides a robust and accurate solution to the initialization of 3D fetal brain reconstruction.

\section*{Acknowledgements}
This research was supported by NIH U01HD087211, R01EB01733, HD100009, NIBIB R01EB032708, NIBIB NAC P41EB015902, R01AG070988, RF1MH123195, ERC Starting Grant 677697, ARUK-IRG2019A-003.

%
%
%
\bibliographystyle{splncs04}
\bibliography{ref.bib}

\begin{thebibliography}{10}
\providecommand{\url}[1]{\texttt{#1}}
\providecommand{\urlprefix}{URL }
\providecommand{\doi}[1]{https://doi.org/#1}

\bibitem{ba2016layer}
Ba, J.L., Kiros, J.R., Hinton, G.E.: Layer normalization. arXiv preprint
  arXiv:1607.06450  (2016)

\bibitem{benkarim2020novel}
Benkarim, O., Piella, G., Rekik, I., Hahner, N., Eixarch, E., Shen, D., Li, G.,
  Ballester, M.A.G., Sanroma, G.: A novel approach to multiple anatomical shape
  analysis: application to fetal ventriculomegaly. Medical image analysis
  \textbf{64},  101750 (2020)

\bibitem{devlin2019bert}
Devlin, J., Chang, M.W., Lee, K., Toutanova, K.: Bert: Pre-training of deep
  bidirectional transformers for language understanding (2019)

\bibitem{dosovitskiy2020image}
Dosovitskiy, A., Beyer, L., Kolesnikov, A., Weissenborn, D., Zhai, X.,
  Unterthiner, T., Dehghani, M., Minderer, M., Heigold, G., Gelly, S., et~al.:
  An image is worth 16x16 words: Transformers for image recognition at scale.
  arXiv preprint arXiv:2010.11929  (2020)

\bibitem{esteban2019towards}
Esteban, J., Grimm, M., Unberath, M., Zahnd, G., Navab, N.: Towards fully
  automatic {X-ray} to {CT} registration. In: International Conference on
  Medical Image Computing and Computer-Assisted Intervention. pp. 631--639.
  Springer (2019)

\bibitem{gagoski2021automated}
Gagoski, B., Xu, J., Wighton, P., Tisdall, M.D., Frost, R., Lo, W.C., Golland,
  P., van Der~Kouwe, A., Adalsteinsson, E., Grant, P.E.: Automated detection
  and reacquisition of motion-degraded images in fetal haste imaging at 3 t.
  Magnetic Resonance in Medicine  (2021)

\bibitem{gholipour2010robust}
Gholipour, A., Estroff, J.A., Warfield, S.K.: Robust super-resolution volume
  reconstruction from slice acquisitions: application to fetal brain {MRI}.
  IEEE transactions on medical imaging  \textbf{29}(10),  1739--1758 (2010)

\bibitem{gholipour2017normative}
Gholipour, A., Rollins, C.K., Velasco-Annis, C., Ouaalam, A., Akhondi-Asl, A.,
  Afacan, O., Ortinau, C.M., Clancy, S., Limperopoulos, C., Yang, E., et~al.: A
  normative spatiotemporal {MRI} atlas of the fetal brain for automatic
  segmentation and analysis of early brain growth. Scientific reports
  \textbf{7}(1),  1--13 (2017)

\bibitem{gillies2017real}
Gillies, D.J., Gardi, L., De~Silva, T., Zhao, S.r., Fenster, A.: Real-time
  registration of 3d to 2d ultrasound images for image-guided prostate biopsy.
  Medical physics  \textbf{44}(9),  4708--4723 (2017)

\bibitem{he2016deep}
He, K., Zhang, X., Ren, S., Sun, J.: Deep residual learning for image
  recognition. In: Proceedings of the IEEE conference on computer vision and
  pattern recognition. pp. 770--778 (2016)

\bibitem{hou20183}
Hou, B., Khanal, B., Alansary, A., McDonagh, S., Davidson, A., Rutherford, M.,
  Hajnal, J.V., Rueckert, D., Glocker, B., Kainz, B.: 3-d reconstruction in
  canonical co-ordinate space from arbitrarily oriented 2-d images. IEEE
  transactions on medical imaging  \textbf{37}(8),  1737--1750 (2018)

\bibitem{hou2018computing}
Hou, B., Miolane, N., Khanal, B., Lee, M.C., Alansary, A., McDonagh, S.,
  Hajnal, J.V., Rueckert, D., Glocker, B., Kainz, B.: Computing cnn loss and
  gradients for pose estimation with riemannian geometry. In: International
  Conference on Medical Image Computing and Computer-Assisted Intervention. pp.
  756--764. Springer (2018)

\bibitem{iglesias2021joint}
Iglesias, J.E., Billot, B., Balbastre, Y., Tabari, A., Conklin, J.,
  Gonz{\'a}lez, R.G., Alexander, D.C., Golland, P., et~al.: Joint
  super-resolution and synthesis of 1 mm isotropic mp-rage volumes from
  clinical {MRI} exams with scans of different orientation, resolution and
  contrast. NeuroImage  \textbf{237},  118206 (2021)

\bibitem{kainz2015fast}
Kainz, B., Steinberger, M., Wein, W., Kuklisova-Murgasova, M., Malamateniou,
  C., Keraudren, K., Torsney-Weir, T., Rutherford, M., Aljabar, P., Hajnal,
  J.V., et~al.: Fast volume reconstruction from motion corrupted stacks of 2d
  slices. IEEE transactions on medical imaging  \textbf{34}(9),  1901--1913
  (2015)

\bibitem{kendall2015posenet}
Kendall, A., Grimes, M., Cipolla, R.: Posenet: A convolutional network for
  real-time 6-dof camera relocalization. In: Proceedings of the IEEE
  international conference on computer vision. pp. 2938--2946 (2015)

\bibitem{kingma2017adam}
Kingma, D.P., Ba, J.: Adam: A method for stochastic optimization (2017)

\bibitem{krishnamurthy2015mr}
Krishnamurthy, U., Neelavalli, J., Mody, S., Yeo, L., Jella, P.K., Saleem, S.,
  Korzeniewski, S.J., Cabrera, M.D., Ehterami, S., Bahado-Singh, R.O., et~al.:
  Mr imaging of the fetal brain at 1.5 t and 3.0 t field strengths: comparing
  specific absorption rate (sar) and image quality. Journal of perinatal
  medicine  \textbf{43}(2),  209--220 (2015)

\bibitem{kuklisova2012reconstruction}
Kuklisova-Murgasova, M., Quaghebeur, G., Rutherford, M.A., Hajnal, J.V.,
  Schnabel, J.A.: Reconstruction of fetal brain {MRI} with intensity matching
  and complete outlier removal. Medical image analysis  \textbf{16}(8),
  1550--1564 (2012)

\bibitem{malamateniou2013motion}
Malamateniou, C., Malik, S., Counsell, S., Allsop, J., McGuinness, A., Hayat,
  T., Broadhouse, K., Nunes, R., Ederies, A., Hajnal, J., et~al.:
  Motion-compensation techniques in neonatal and fetal mr imaging. American
  Journal of Neuroradiology  \textbf{34}(6),  1124--1136 (2013)

\bibitem{paszke2017automatic}
Paszke, A., Gross, S., Chintala, S., Chanan, G., Yang, E., DeVito, Z., Lin, Z.,
  Desmaison, A., Antiga, L., Lerer, A.: Automatic differentiation in pytorch
  (2017)

\bibitem{payette2021automatic}
Payette, K., de~Dumast, P., Kebiri, H., Ezhov, I., Paetzold, J.C., Shit, S.,
  Iqbal, A., et~al.: An automatic multi-tissue human fetal brain segmentation
  benchmark using the fetal tissue annotation dataset. Scientific Data
  \textbf{8}(1),  1--14 (2021)

\bibitem{pei2020anatomy}
Pei, Y., Wang, L., Zhao, F., Zhong, T., Liao, L., Shen, D., Li, G.:
  Anatomy-guided convolutional neural network for motion correction in fetal
  brain {MRI}. In: International Workshop on Machine Learning in Medical
  Imaging. pp. 384--393. Springer (2020)

\bibitem{salehi2018real}
Salehi, S.S.M., Khan, S., Erdogmus, D., Gholipour, A.: Real-time deep pose
  estimation with geodesic loss for image-to-template rigid registration. IEEE
  transactions on medical imaging  \textbf{38}(2),  470--481 (2018)

\bibitem{simonyan2014very}
Simonyan, K., Zisserman, A.: Very deep convolutional networks for large-scale
  image recognition. arXiv preprint arXiv:1409.1556  (2014)

\bibitem{singh2020deep}
Singh, A., Salehi, S.S.M., Gholipour, A.: Deep predictive motion tracking in
  magnetic resonance imaging: application to fetal imaging. IEEE Transactions
  on Medical Imaging  \textbf{39}(11),  3523--3534 (2020)

\bibitem{tustison2010n4itk}
Tustison, N.J., Avants, B.B., Cook, P.A., Zheng, Y., Egan, A., Yushkevich,
  P.A., Gee, J.C.: N4itk: improved n3 bias correction. IEEE transactions on
  medical imaging  \textbf{29}(6),  1310--1320 (2010)

\bibitem{vasung2019exploring}
Vasung, L., Turk, E.A., Ferradal, S.L., Sutin, J., Stout, J.N., Ahtam, B., Lin,
  P.Y., Grant, P.E.: Exploring early human brain development with structural
  and physiological neuroimaging. Neuroimage  \textbf{187},  226--254 (2019)

\bibitem{vaswani2017attention}
Vaswani, A., Shazeer, N., Parmar, N., Uszkoreit, J., Jones, L., Gomez, A.N.,
  Kaiser, {\L}., Polosukhin, I.: Attention is all you need. Advances in neural
  information processing systems  \textbf{30} (2017)

\bibitem{wang2004image}
Wang, Z., Bovik, A.C., Sheikh, H.R., Simoncelli, E.P.: Image quality
  assessment: from error visibility to structural similarity. IEEE transactions
  on image processing  \textbf{13}(4),  600--612 (2004)

\bibitem{xu2021stress}
Xu, J., Abaci~Turk, E., Grant, P.E., Golland, P., Adalsteinsson, E.: Stress:
  Super-resolution for dynamic fetal {MRI} using self-supervised learning. In:
  International Conference on Medical Image Computing and Computer-Assisted
  Intervention. pp. 197--206. Springer (2021)

\bibitem{yeung2021learning}
Yeung, P.H., Aliasi, M., Papageorghiou, A.T., Haak, M., Xie, W., Namburete,
  A.I.: Learning to map 2d ultrasound images into 3d space with minimal human
  annotation. Medical Image Analysis  \textbf{70},  101998 (2021)

\end{thebibliography}

\end{document}